\documentclass{raa}           
\usepackage{graphicx,times}
\usepackage{natbib}
\usepackage{amssymb,amsmath}
\usepackage{mathrsfs}
\bibpunct{(}{)}{;}{a}{}{,}

\usepackage[a4paper=true,dvipdfm=true,pagebackref=true]{hyperref}
\hypersetup{pdftitle = The title of my PDF, pdfauthor = My name, pdfsubject= The subject, pdfkeywords = keyword1 keyword2 keyword3} 
\hypersetup{colorlinks = true, linkcolor = green, anchorcolor = red, citecolor = blue, filecolor = red, pagecolor = red, urlcolor = red}

\begin{document}
	
	\title{The influence of the \emph{Insight}-HXMT/LE time response on timing analysis
	}

	\volnopage{ {\bf 20XX} Vol.\ {\bf X} No. {\bf XX}, 000--000}
	\setcounter{page}{1}

	\author{Deng-Ke Zhou\inst{1,2}, Shi-Jie Zheng\inst{1}, Li-Ming Song$^*$\inst{1,2}, Yong Chen\inst{1}, Cheng-Kui Li\inst{1}, Xiao-Bo Li\inst{1}, Tian-Xiang Chen\inst{1}, Wei-Wei Cui\inst{1}, Wei Chen\inst{1,2}, Da-Wei Han\inst{1}, Wei Hu\inst{1}, Jia Huo\inst{1}, Rui-Can Ma\inst{1,2}, Mao-Shun Li\inst{1}, Tian-Ming Li\inst{1,2}, Wei Li\inst{1}, He-Xin Liu\inst{1,2}, Bo Lu\inst{1}, Fang-Jun Lu\inst{1}, Jin-Lu Qu\inst{1,2}, You-Li Tuo\inst{1,2}, Juan Wang\inst{1}, Yu-Sa Wang\inst{1}, Bai-Yang Wu\inst{1,2}, Guang-Cheng Xiao\inst{1,2}, Yu-Peng Xu\inst{1}, Yan-Ji Yang\inst{1}, Shu Zhang\inst{1}, Zi-Liang Zhang\inst{1}, Xiao-Fan Zhao\inst{1,2}, Yu-Xuan Zhu\inst{1,3}}
	
	\institute{ Key Laboratory of Particle Astrophysics, Institute of High Energy Physics, Chinese Academy of Sciences, 19B Yuquan Road, Beijing 100049, China; {\it songlm@ihep.ac.cn, zhoudk@ihep.ac.cn}\\
		\and
		University of Chinese Academy of Sciences, Chinese Academy of Sciences, Beijing 100049, China\\
		\and 
		College of Physics, Jilin University, 2699 Qianjin Street, Changchun City, 130012, China\\
		\vs \no
		{\small Received 20XX Month Day; accepted 20XX Month Day}
	}
	
	\abstract{LE is the low energy telescope of \emph{Insight}-HXMT. It uses swept charge devices (SCDs) to detect soft X-ray photons. The time response of LE is caused by the structure of SCDs. With theoretical analysis and Monte Carlo simulations we discuss the influence of LE time response (LTR) on the timing analysis from three aspects: the power spectral density, the pulse profile and the time lag. After the LTR, the value of power spectral density monotonously decreases with the increasing frequency. The power spectral density of a sinusoidal signal reduces by a half at frequency 536 Hz. The corresponding frequency for QPO signals is 458 Hz. The Root mean square (RMS) of QPOs holds the similar behaviour. After the LTR, the centroid frequency and full width at half maxima (FWHM) of QPOs signals do not change. The LTR reduces the RMS of pulse profiles and shifts the pulse phase. In the time domain, the LTR only reduces the peak value of the cross-correlation function while it does not change the peak position. Thus it will not affect the result of the time lag. 
		When considering the time lag obtained from two instruments and one among them is LE, a 1.18 ms lag is expected caused by the LTR.
		The time lag calculated in the frequency domain is the same as that in the time domain. 
	}
	
	\authorrunning{D. K. Zhou, S. J. Zheng and L. M. Song et al.}            
	\titlerunning{The influence of the \emph{Insight}-HXMT/LE time response on timing analysis}  
	\maketitle

	%
	\section{Introduction}
\begin{figure}[h!]
		\centering
		\includegraphics[scale=.6]{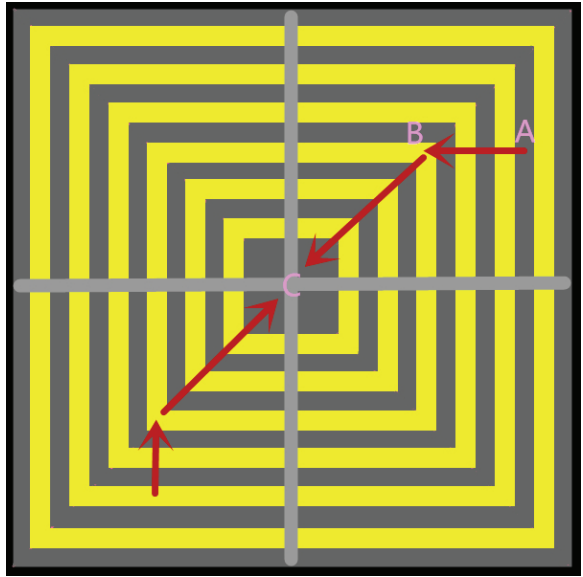}
		\caption{Schematic diagram of the charge collection by LE/SCDs. Each SCD has four quadrants. In each quadrant the yellow and gray areas are called 'packed-pixles'. The color is just for visual discrimination. There are 120 'packed-pixels' in each quadrant, but we only draw a few of them in the diagram. The charge generated by the 'packed-pixel' A is first transferred to B, and then transferred to the central C for recording. The incident photons at different positions of the SCD lead to different charge collection time, i.e., the time response.}
		\label{figure_detector}
	\end{figure}

	The Hard X-ray Modulation Telescope launched on $15^{th}$ June 2017, also dubbed as \emph{Insight}-HXMT, is China's first X-ray astronomical satellite (\citealt{Intro, Zhang+etal+2020}). It carries three payloads: High Energy X-ray Telescope (HE), Medium Energy X-ray Telescope (ME) and Low Energy X-ray Telescope (LE). The LE contains three identical detector boxes (LEDs) and one electric control box (LEB). Each LED consists of eight swept charge device (SCD) modules, two types of collimators, visible light blocking filters, anti-contamination films, heat pipes as well as several thermal and mechanical supporters (\citealt{LE_consists}). The LE uses SCDs to record time-of-arrivals and energies of soft X-ray photons in the energy band of 0.7-13\,keV. The time response of the LE is caused by the structure of SCDs. There are four quadrants in each SCD detector. The charges generated by incident photons that hit the detector flow out along a specific path in each quadrant. The readout time of photons hitting at different positions of the detector is different, which results in the time response distribution (TRD) (\citealt{Intro_zhao}). Fig \ref{figure_detector} is a schematic diagram of the charge collection of each LE/SCD.
	
	Assuming a triangular TRD distribution inferred from the structure of LE/SCDs, 
	\citet{Method_conv} studied the time response of LE and its influence on the pulse profile of pulsations. 
	After that, \citet{Intro_zhao} measured the TRD of LE/SCDs using a long exposure readout mode (LERM) and obtained the probability distribution of the readout time. In this paper, we perform a more reasonable analysis using the updated TRD.
	
	The structure of this paper is as follows: in section \ref{section_method}, the influence of the LE time response (hereafter LTR) on the intensity function is studied using analytical methods. We also investigated the LTR effect using Monte Carlo simulations. In section \ref{section_PSD}, the influence of the LTR on the power spectral density (PSD) of sinusoidal signals and quasi-periodic oscillations (QPOs) is discussed. In section \ref{section_OnPulseProfile}, the influence on the pulse profile of pulsar signals is discussed. Section \ref{section_OnTimeLag} focuses on the time lag and section \ref{section_conclusion} gives the discussion and conclusions of the paper.
	\section{Methodology}
	\label{section_method}
	\subsection{Theoretical Analysis}
	\label{subsection_analysis}
	The number of detected X-ray photons ($N$) in a given time interval $(t_{min}, t_{max})$ obeys a Poisson distribution with a parameter $\lambda(t)$ (\citealt{Method_NP2}), i.e., 
	\begin{equation}
		P(N=k)=\frac{\left[\int_{t_{min}}^{t_{max}} \lambda(t) dt\right]^{k} exp\left[-\int_{t_{min}}^{t_{max}} \lambda(t) dt\right]}{k!}, 
		\label{equation_possion}
	\end{equation}
	where $P(N=k)$ represents the possibility of detecting $k$ photons in the given time interval $(t_{min}, t_{max})$; $\lambda(t)$\ is called the time-varying rate or the intensity function in the unit of photons per second and $\int_{t_{min}}^{t_{max}} \lambda(t) dt=E[N]$, where $E[N]$ is the expectation of the variable $N$. Let $T_0$ represent the actual arrival time of an X-ray photon and $D$ the readout time. Therefore, the time $Z$ recorded by the LE/SCD is that
	\begin{equation}
		\begin{aligned}
			Z=T_0+D,\\
			D=D_0+D_k.
		\end{aligned}
	\end{equation}
	Here the $D$ is composed of $D_0$ and $D_k$, where $D_0$ equals to $10^{-5}$ second, which is the fixed delay caused by the 100 kHz working frequency of the readout mode of LE/SCDs, and $D_k$ is the time delay caused by the TRD effect.
	
\begin{figure}
		\centering
		\includegraphics[scale=.6]{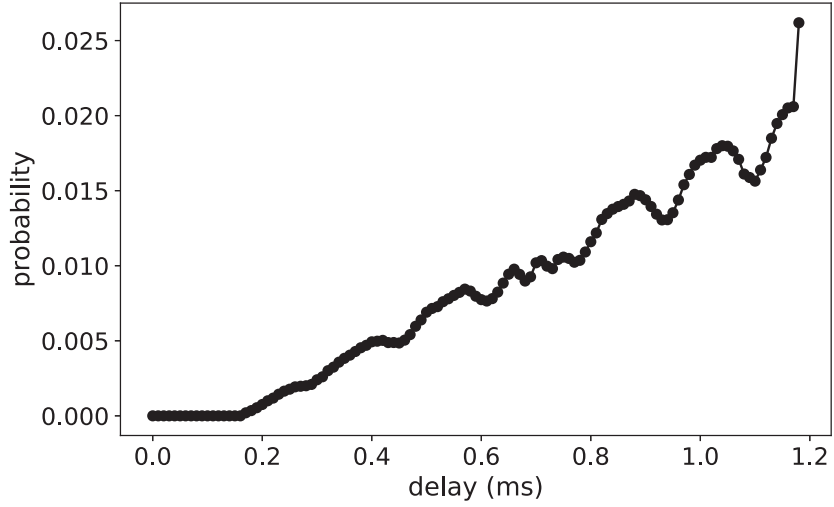}
		\caption{Time response distribution (TRD) of LE/SCD.}
		\label{figure_TRD}
	\end{figure}

	The intensity function recorded by LE/SCDs is influenced by the TRD effect, which can be described as the convolution of the original intensity function $\lambda(t)$ and the TRD (\citealt{Method_conv}). 
	Considering the updated TRD and the fixed delay , the response intensity function $\lambda'(t)$ is given by
	\begin{equation}
		\label{equation_main}
		\begin{aligned}
			\lambda'(t)=\lambda(t)*h(t),  \\
			h(t)=\sum_{k=1}^{n}p_k\delta[t-10^{-5}(k+1)], \\
			\sum_{k=1}^{n}p_k=1, 
		\end{aligned}
	\end{equation}
	where n=118 and $\lambda(t)$ is the original intensity function; $p_k$ is the $ kth$ probability of delaying $10^{-5}k$ seconds. $p_k$ and the corresponding delay are shown in Fig. \ref{figure_TRD}; $\delta$ is the Dirac delta function. Thus $h(t)$ is the impulse response function of LE/SCDs. The detection area of each SCD quadrant is a L-shaped strip, which is called the 'packed-pixel' (Fig. \ref{figure_detector}). There are 120 'packed-pixels' in each quadrant. The 20 'packed-pixels' near the centre have a fixed read-out time. On the other hand, two 'packed-pixels' close to the outskirt of SCDs are invalid, which are thus not considered. Therefore, with the working frequency of 100 kHz, all charges would
	be readout within 1.18 ms (\citealt{Intro_zhao}). Because the 20 'packed-pixels' near the centre have a fixed delay and the area is smaller than the outside, the time resolution of LE/SCDs is often recorded as 0.98 ms (\citealt{LE_consists}). In this paper, all delay effects are taken into account. So the sequence number of formula (\ref{equation_main}) starts from 1 to 118. The LTR effect in the timing analysis can be studied by comparing $\lambda'(t)$ with $\lambda(t)$. 
	\subsection{Monte Carlo simulation}\label{subsection_MC}
	The Monte Carlo (MC) simulation is widely used when dealing with difficult quantitative analysis. The general method is as follows:
	\begin{enumerate}
		\item In the time interval $(t_{min}, t_{max})$, the time set $\{t_i\}_{i=1}^{M}$ that represents arrival times of photons is sampled from the original intensity function $\lambda(t)$ by using the rejection sampling method. The time set $\{t_i\}_{i=1}^{M}$ satisfies $t_1\leq t_2\leq...\leq t_i\leq...\leq t_{M-1}\leq t_M$.The time set $\{t_i\}_{i=1}^{M}$ follows a given distribution, for examples, periodic signals, QPOs signals, etc. 
		\item Add a random number to each value in the time set $\{t_i\}_{i=1}^{M}$ to get a new time set $\{u_i\}_{t=1}^M$, where $u_i=t_i+D$. We arrange $\{u_i\}_{t=1}^M$ in a ascending order and still denote it as $\{u_i\}_{t=1}^M$. Therefore, $\{u_i\}_{t=1}^M$ satisfies $t_1+10^{-5}\leq u_1\leq u_2\leq...\leq u_i\leq...\leq u_{M-1}\leq u_M \leq t_M+118\times 10^{-5}$. 
		\item Extract two light curves from time sets $\{t_i\}_{i=1}^{M}$ and  $\{u_i\}_{t=1}^M$, which represent signals before and after the LTR, respectively. Then the influence of the LTR on the timing analysis can be studied by comparing these two light curves. 
	\end{enumerate}
	
	In order to carry out the simulation more efficiently, a spectral-timing software package \emph{stingray} for astrophysical X-ray data analysis is employed (\citealt{stingray}).  
	
	\section{On power spectral density (PSD)}
	\label{section_PSD}
	\subsection{General analysis}
	\label{subsection_sinPower}
\begin{figure}
		\includegraphics[scale=0.8]{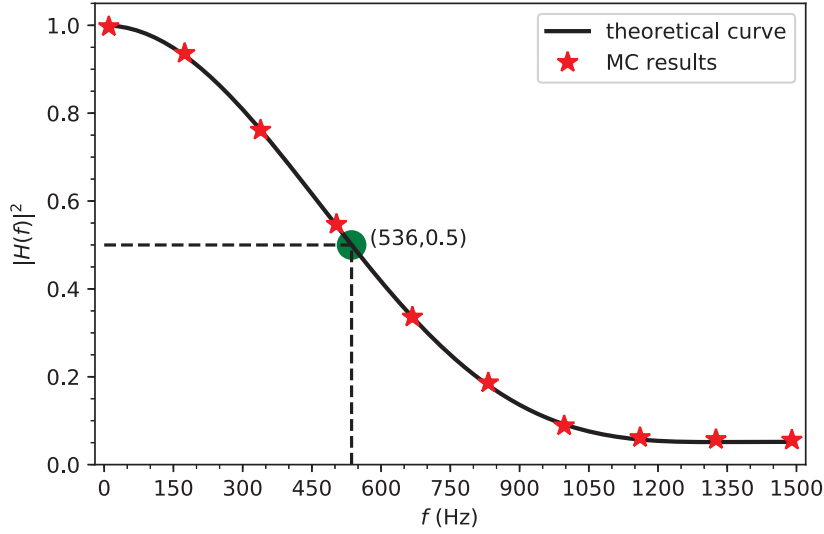}
		\centering
		\caption{$|H(f)|^2$ as a function of frequency. The solid line is $|H(f)|^2$ and the star markers are results of Monte Carlo simulations. The green dot indicates that the PSD peak after the LTR at 536 Hz decreases by a half.}
		\label{figure_Hf}
	\end{figure}

	Let $P_\lambda(f)$ and $P_{\lambda'}(f)$ represent PSDs of $\lambda(t)$ and $\lambda'(t)$, respectively. Thus, their relation can be written as:
	\begin{equation}
		P_{\lambda'}(f)=|H(f)|^2P_{\lambda}(f), 
	\end{equation}
	where $H(f)$ is the Fourier transform of the impulse response function $h(t)$ and $f$ represents the frequency. Considering the formula (\ref{equation_main}), $|H(f)|^2$ is given by
	\begin{equation}
		\label{equation_Hf}
		\begin{aligned}
			|H(f)|^2={} & \mathscr{F}[h(t)]\mathscr{F}[h(t)]^*\\
			={} & \sum_{k=1}^{n}p_k^2+2\sum_{i=1}^{n-1}\sum_{j>i}^{n}p_ip_j{\rm cos}[2\times 10^{-5}\pi f(i-j)]. 
		\end{aligned}
	\end{equation}
	Obviously, 
	\begin{equation}
		|H(f)|^2<(\sum_{k=1}^{n}p_k^2+2\sum_{i=1}^{n-1}\sum_{j>i}^{n}p_ip_j)=(\sum_{k=1}^{n}p_k)^2=1, 
	\end{equation}
	where $n$=118. Therefore, the $P_{\lambda'}(f)$ is smaller than $P_{\lambda}(f)$ at any frequency and the attenuation factor is $|H(f)^2|$ (shown in Fig. \ref{figure_Hf}).

	In addition, we performed MC simulations using the method mentioned above. 
	In practice, we assumed that the original light curve has an intensity of 1000\,cts/s and an exposure of 100\,s. 
	We show the modulation of the PSD caused by the LTR in Fig.~\ref{figure_Hf}.
	We found that our simulations are well consistent with the theoretical estimation.
	
	\subsection{QPO signals}
	\label{subsection_QPOs}
	We used the MC method to simulate quasi-periodic oscillations (QPOs), which are typical signals in the astronomical timing analysis. In the PSD, QPOs commonly appear as a peak at a certain frequency with a finite width, which can be described as a Lorentzian function \citep{QPO_lorentzian}
	\begin{equation}
		P_\nu = \frac{A_0\omega}{(\nu-\nu_0)^2+(\frac{\omega}{2})^2}
		\label{equation_lorenzian}, 
	\end{equation}
	where $\nu_0$ is the centroid frequency, $\omega$ is the full-width at half maximum (FWHM), and $A_0$ is the amplitude of the signal. The quality factor (define Q$\equiv \frac{\nu_0}{\omega}$) represents the significance of QPOs. By convention, signals with $Q>2$ are called QPOs while those with $Q\leq 2$ called the peaked noise (\citealt{QPO_lorentzian}). 
	
	We simulated different centroid frequencies of QPOs. The detailed process is as follows:
	\begin{itemize}
		\item Consider Lorentzian-shape PSDs described by the formula (\ref{equation_lorenzian}) with specific parameters.
		In practice, we performed simulations assuming variable centroid frequencies between 10\,Hz and 600\,Hz with a resolution of 10\,Hz.
		We also set $A_0\equiv1$ and $\omega\equiv\frac{\nu_0}{10}$. Therefore, the Q factor of QPOs signals we simulated always equals to 10.
		\item Simulate a light curve for a given PSD using the algorithm proposed by \citet{QPO_psd2lc}. The light curve duration and root mean square (RMS) were set at 500 s and 0.5, respectively. The time resolution of the light curve was set at 1/4 reciprocal of the centroid frequency of the corresponding QPOs.   
		\item Using the light curve as a intensity function, the time set $\{t_i\}_{i=1}^M$ before the LTR and the time set $\{u_i\}_{i=1}^M$ after the LTR were obtained by using the method mentioned in subsection \ref{subsection_MC}. 
		\item Generate PSDs using the time sets $\{t_i\}_{i=1}^M$ and $\{u_i\}_{i=1}^M$. 
		In order to suppress the Poisson noise, each light curve was divided into 10-second segments to calculate PSDs independently, and then an averaged PSD could be obtained. 
		
		Finally, we got resulting parameters of QPOs after the LTR, and compared them with the input models. 
		
	\end{itemize}
	
	\begin{figure}
		\centering
		\includegraphics[scale=0.4]{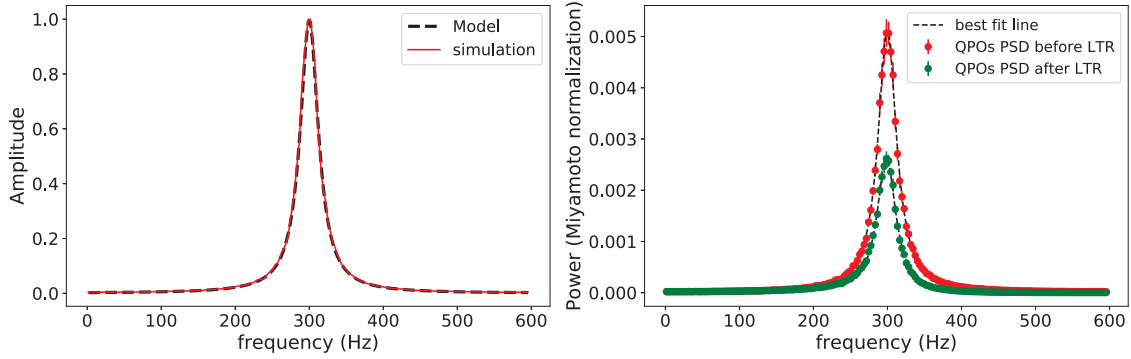}
		\caption{Left panel:The black dashed line is a Lorenzian model and the red line is the simulated PSD which has been normalized for comparison. Right panel:the simulated PSD by using the Lorenzian model. Red points are simulated PSD before the LTR while green points are simulated PSD after the LTR. The black dashed line is the best-fitting with the corresponding Lorenzian model.}
		\label{figure_QPO}
	\end{figure}
	
	Fig. $\ref{figure_QPO}$ shows an example. The PSD calculated here is based on the Miyamoto normalization, which is a convenient method for calculating the RMS of QPOs  (\citealt{QPO_PSDnorm_miya}). 
	
\begin{figure*}
		\includegraphics[width=\textwidth]{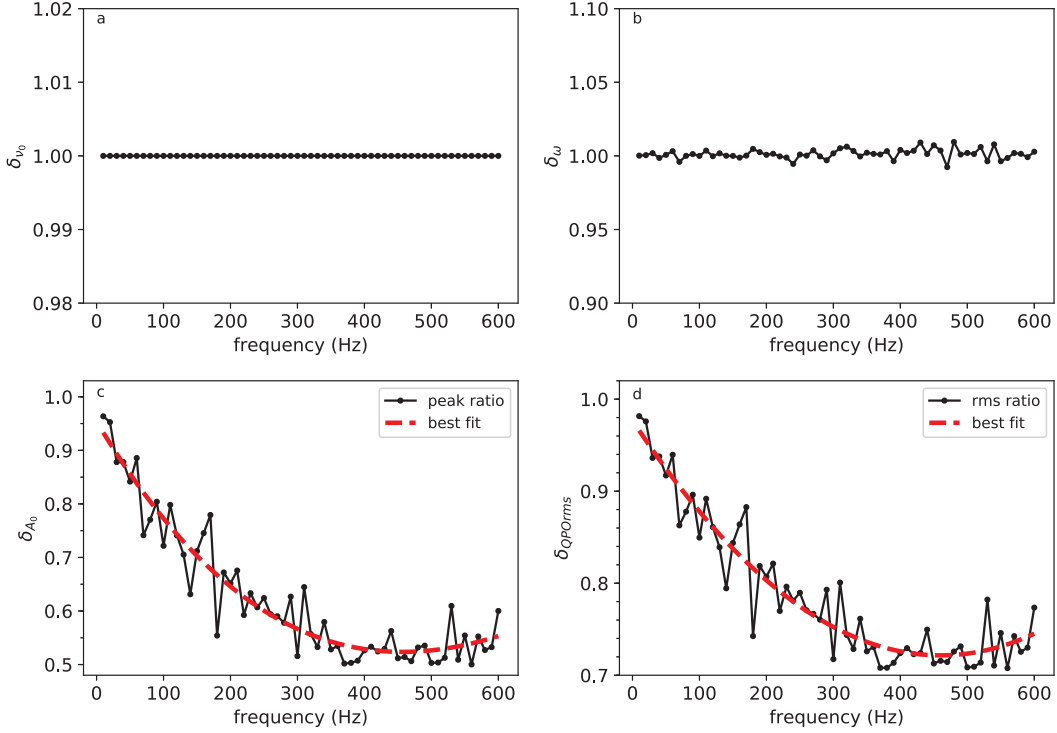}
		\centering
		\caption{ 
			The influence of the LTR effect on the QPO studies.
			From a practical point of view, the higher frequencies have a larger simulation error so we only got to 600 Hz. Panel a: $\delta_{\nu_0}$ as the function of frequency. Panel b: $\delta_{\omega}$ as the function of frequency. Panel c: $\delta_{A_0}$ as the function of frequency. The red dashed line is the result of the quadratic polynomial fitting. The fitting function is $y=ax^3+bx^2+cx+d$ and the fitting parameters are $a=-1.05\times 10^{-9}$, $b=3.02\times 10^{-6}$, $c=-2.10\times 10^{-3}$, $d=0.954$, respectively. Panel d: $\delta_{QPOrms}$ as the function of frequency. The red dashed line is the result of the quadratic polynomial fitting. The fitting function is $y=ax^2+bx+c$ and the fitting parameters are $a=1.21\times 10^{-6}$, $b=-1.11\times 10^{-3}$, $c=0.977$, respectively.}
		\label{figure_QPO_results}
	\end{figure*}

	The influence of the LTR on QPOs parameters can be obtained as a function of the QPO frequency, which is defined as $\delta \equiv \frac{\beta}{\alpha}$. 
	$\alpha$ represents the parameters of QPOs before the LTR, such as $\nu_0$, $\omega$, $A_0$ and RMS, and $\beta$ represents the corresponding values after the LTR. The $\delta$ is their ratio, which is denoted as $\delta_{\nu_0}$, $\delta_{\omega}$, $\delta_{A_0}$, $\delta_{QPOrms}$, respectively. Fig. \ref{figure_QPO_results} shows the changes of $\delta_{\nu_0}$, $\delta_{\omega}$, $\delta_{A_0}$ and $\delta_{QPOrms}$ with the QPO frequency.  $\delta_{\nu_0}$ and $\delta_\omega$ are almost invariant, which indicates that they are not affected by the LTR.
	
	On the other hand, $\delta_{A_0}$ of QPOs decreases with the increasing frequency, indicating that the LTR has a great influence on the peak value of QPOs. This is similar to the results obtained from subsection \ref{subsection_sinPower}. We used a quadratic polynomial function $y=ax^3+bx^2+cx+d$ to fit $\delta_{A_0}$. The fitting parameters are $a=-1.05\times 10^{-9}$, $b=3.02\times 10^{-6}$, $c=-2.10\times 10^{-3}$, $d=0.954$, respectively. According to the simulation, the Q factor of QPOs does not change with the centroid frequency because both the FWHM $\omega$ and the centroid frequency $\nu_0$ of QPOs are unchanged after the LTR. Considering that the peak value of QPOs decreases as the frequency increasing, the actual data contains high frequency QPOs may miscalculate the Q factor because the significance of QPOs obtained is too low. The RMS results are shown in the panel d of Fig \ref{figure_QPO_results}. It is similar to $\delta_{A_0}$, which decreases with the increase of the centroid frequency. We fitted the results with a function $y=ax^2+bx+c$. The fitting parameters are $a=1.21\times 10^{-6}$, $b=-1.11\times 10^{-3}$, $c=0.977$, respectively.
	
	\section{On pulse profile of periodic signal}
	\label{section_OnPulseProfile}
	The conventional method to obtain the pulse profile of a pulsar signal is epoch-folding (\citealt{PSR_Ge2012}). Considering that the LTR changes the arrival time of photons in milliseconds, the accuracy of the pulse profile with a period of milliseconds is expected to be influenced significantly. 
	
\begin{figure*}
		\includegraphics[scale=0.45]{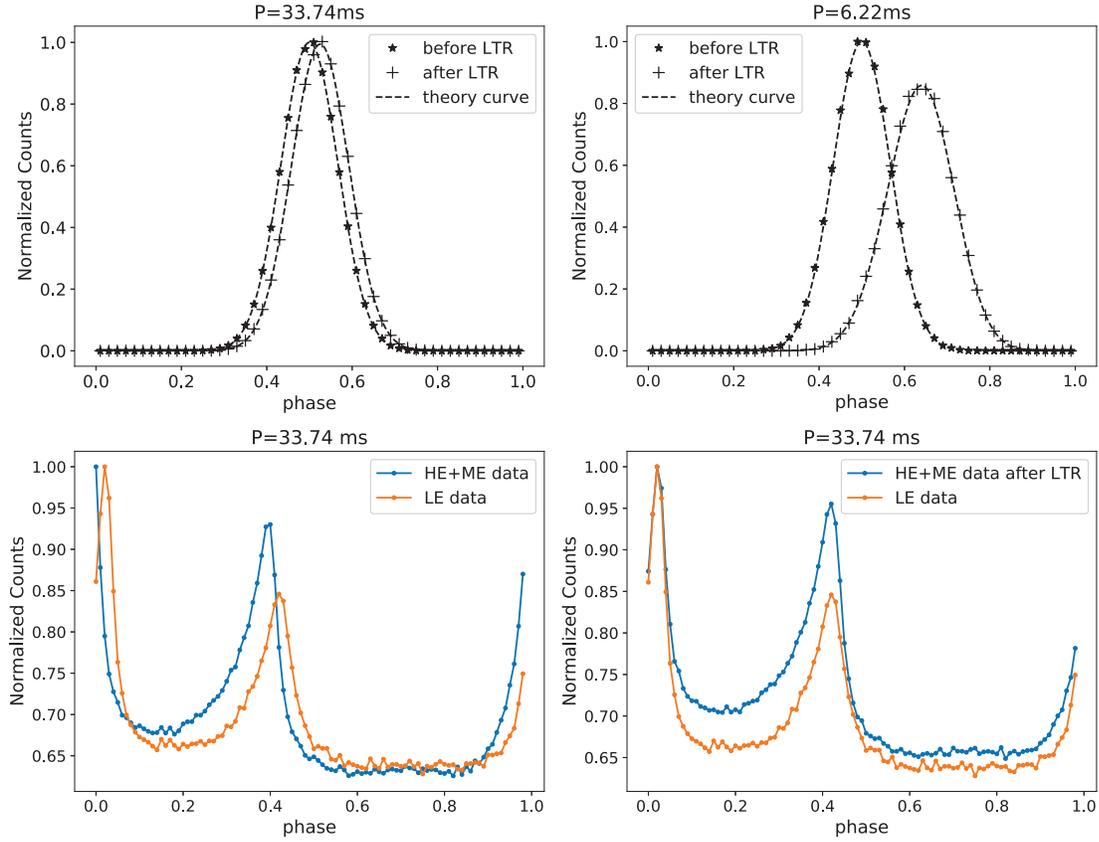}
		\centering
		\caption{Upper left and right panels: intensity functions $\lambda(t)$ and $\lambda'(t)$ before (star sign) and after (cross sign) the LTR assuming different periods.The profiles are obtained by folding the time sets generated by the Monte Carlo method mentioned in subsection \ref{subsection_MC}, and the dashed line is the theoretical prediction by using the formula (\ref{equation_main}). Bottom left panel: the pulse profile of the Crab pulsar observed by \emph{Insight}-HXMT satellite. Bottom right panel: the pulse profiles of HE and ME after the LTR are compared with that of LE.}
		\label{figure_profile}
	\end{figure*}

\begin{figure*}
		\includegraphics[scale=0.43]{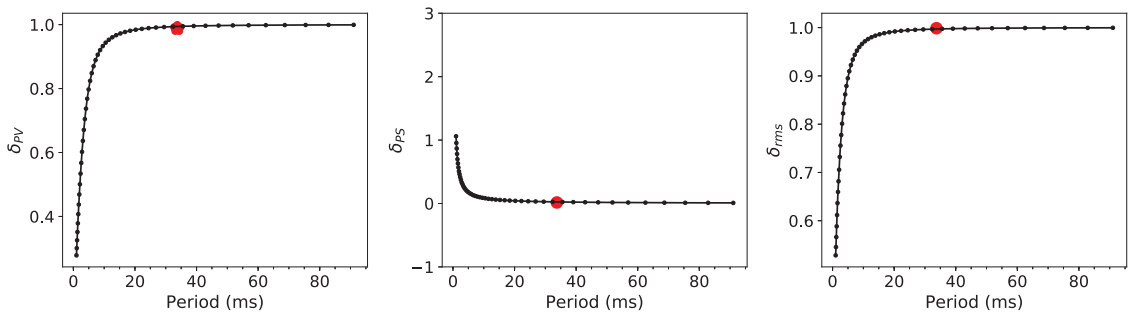}
		\centering
		\caption{$\delta_{PV}$, $\delta_{PS}$ and $\delta_{rms}$ as a function of the pulsed period.  
			The red dots are the results using the Crab data obtained from HE and ME after the LTR.}
		\label{figure_deltaprofile}
	\end{figure*}

	In fact, the pulse profile with a high SNR can be considered as the intensity function mentioned in subsection \ref{subsection_analysis}. The theoretical results are also represented by formula (\ref{equation_main}). Actually, the impulse response function has described the influence on profile already, but in order to get a more intuitive result we simulate a periodic signal. In other words, suppose that the intensity function in one cycle has the following form:
	\begin{equation}
		\lambda(t)=exp[-\frac{(t-\mu)^2}{2\delta ^2}], 
	\end{equation}
	where $\mu$ and $\delta$ represent the peak position and the dispersion degree of the profile in a period $P$, respectively. Here we set $\mu=\frac{P}{2}$ and $\delta=\frac{P}{15}$, respectively. In this way, a single peak signal can be well constructed in one cycle. Once we get the expression of the profile $\lambda(t)$, we can calculate the $\lambda'(t)$ using the formula (\ref{equation_main}) or the Monte Carlo method mentioned in subsection \ref{subsection_MC}. Fig. \ref{figure_profile} shows the results of the profile before and after the LTR when the period $P$ takes different values. It can be seen that the smaller the periodic of the signal, the greater the influence of the LTR, which are represented by the phase shifting (PS) and the peak value (PV). 
	A series of signals with different periods are simulated, and changes of the peak value (denote as $\delta_{PV}$), the phase shifting (denote as $\delta_{PS}$) and the RMS (denote as $\delta_{rms}$) of the signals before and after the LTR are calculated to quantitatively describe the influence of the LTR on the pulse profile. The definitions of $\delta_{PV}$ and $\delta_{rms}$ are similar to those mentioned in subsection \ref{subsection_QPOs}. The definition of $\delta_{PS}$ is $\delta_{PS} \equiv \frac{p_a-p_b}{P}$, where $p_a$ and $p_b$ represent the peak position before and after the LTR and $P$ represents the period. 
	Fig. \ref{figure_deltaprofile} shows the simulation results.
	
	In addition, we also used the real data of the \emph{Insight}-HXMT (ObsID: P010129900101) to do some verification. The data reduction and the scientific results have been published by \citet{PSR_crab_data}. We found that the phase of the LE pulse profile is shifted, compared with HE and ME (Fig. \ref{figure_deltaprofile}). For the convenience of comparison, Fig. \ref{figure_deltaprofile} also shows the HE and ME results after the LTR. It appears that the peak position of HE and ME after the LTR is indeed close to the peak position observed by the LE, which indicates that the abnormality of the LE pulse profile is mainly caused by the LTR. For comparison, we draw the $\delta_{PV}$, $\delta_{PS}$ and $\delta_{rms}$ of HE and ME in Fig. \ref{figure_deltaprofile} with red dots, which are in good agreement with the simulated curve.

	\section{On the time lag}
	\label{section_OnTimeLag} 
	\subsection{the the time domain time lag}
	In the time domain, the time lag between two signals is usually calculated using the cross-correlation function (\citealt{timelag1}). The time displacement that maximizes the cross-correlation function is called the time delay between the two signals. In this section, we simulate two signals with a fixed time delay. Then we calculate their time lag after the LTR. In this way, we explore how the LTR influence the time lag. 
	
\begin{figure*}
		\includegraphics[width=\textwidth]{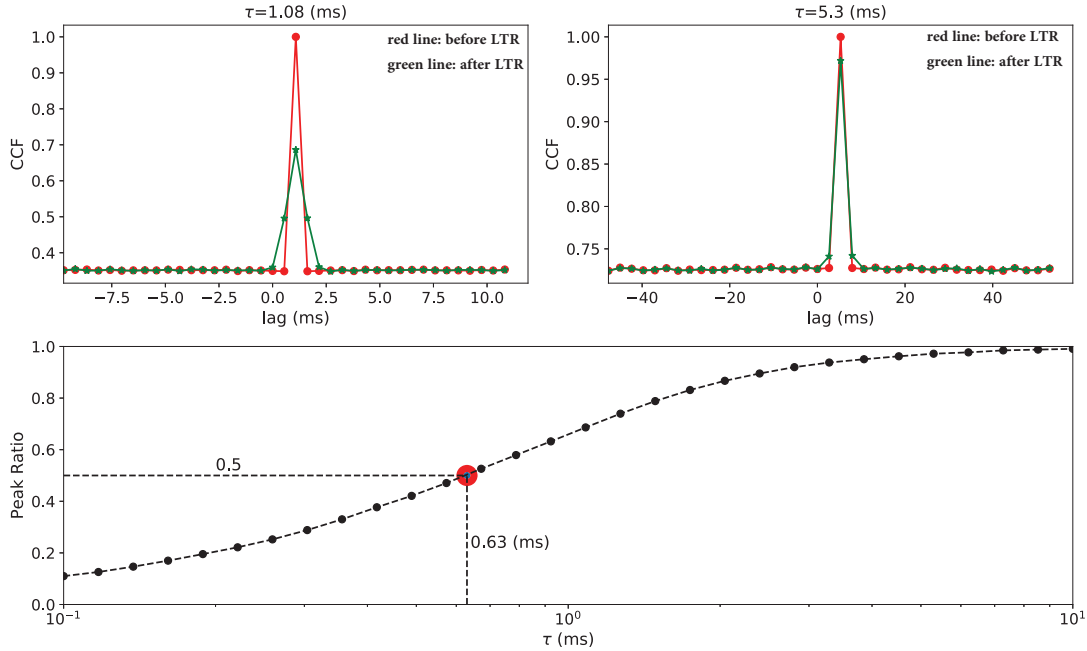}
		\centering
		\caption{Simulations for the time lag in the time domain. Top left and right panels: the cross-correlation functions before (red) and after (green) the LTR, assuming different time lags. 
			Bottom panel: the peak ratio of the cross-correlation function as a function of $\tau$. When the $\tau$ is 0.63 ms, the peak value of cross-correlation function decreases by a half (red dot). However, the peak position does not change, which implies that there is no influence on the time lag detected with \textit{Insight}-HXMT/LE.
		}
		\label{figure_timelag}
	\end{figure*}

	For the sake of simplify, we chose a flat signal as the intensity function $\lambda(t)$. We sampled this intensity function to get the time set which represent arrival times of a batch of photons. Then we shifted the time sets with a constant $\tau$ to get a new time set. These two time sets have a fixed time lag $\tau$. 
	Then cross-correlation function between light curves extracted from these two time sets were calculated, which is shown in Fig. \ref{figure_timelag}. 
	The time resolution of light curves was set at half of the time lag $\tau$.
	Two cross-correlation functions (before and after the LTR) are compared. It is clear that the smaller the time lag, the smaller the peak of the cross-correlation function. The LTR only reduces the values of the function and does not change the maximum position of it, so the calculated time lag actually will not be influenced. 
	
	However, when we consider light curves detected with LE and ME/HE, the LTR effect is expected to have an influence on the lag detection. 
	In this case, two non-lag time sets were generated, one of which was responsed by the LE LTR. After that, the cross-correlation function was calculated between these two signals to obtain the time lag. This simulation shows that the resulting time lag is 1.18 millisecond. It is the maximum time delay of the LTR. This is reasonable because the intensity function is mainly delayed by 1.18 millisecond.
	
	\subsection{the frequency domain time lag}
\begin{figure}
		\includegraphics[scale=.36]{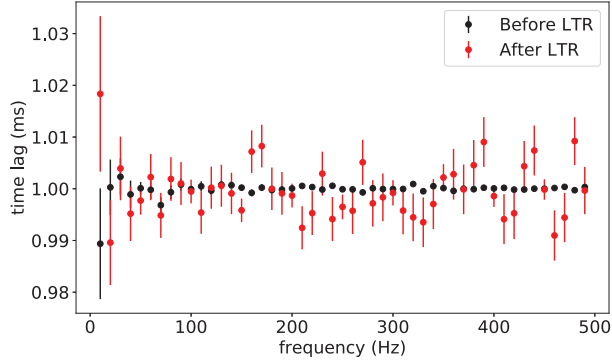}
		\centering
		\caption{Simulation results of the time lag in the frequency domain. Black (red) points represent the time lag before (after) the LTR.
			The assumed time lag is 1 ms in the simulation. After the LTR, the time lag is still close to the true value but more diffuse.}
		\label{figure_timelag2}
	\end{figure}

	We used the cross-spectrum to estimate the time lag of two signals at different frequencies (\citealt{stingray}). Because we want to calculate the signal time lag at different frequencies, we assumed a set of sinusoidal signals to produce the the intensity function, and generated a time set. Then we shifted the time set to get a delayed time set. Finally, we extract two light curves from these two time sets. These two light curves were used to calculate the cross-spectrum to obtain the time lag.  
	The result is similar to that of the cross-correction function. As shown in Fig. \ref{figure_timelag2}, after the LTR the time lag between two signals is not changed, despite the results become more diffuse. The time lag, in case that one signal responsed by LE while the other not, in the frequency domain is also similar to that in the time domain. 
	
	\section{Discussion and conclusion}
	\label{section_conclusion}
	The timing analysis mainly depends on the analysis of light curves. The light curve is the sampling implementation of the intensity function. Therefore we analyzed the influence of the LTR on the intensity function at the beginning of this paper. The effect of the LTR is the convolution of the original intensity function $\lambda(t)$ and the TRD. In addition, the Monte Carlo method is introduced to facilitate the calculation of some quantities. We then discuss the influence of the LTR on the timing analysis from three aspects: the PSD, the pulse profile and the time lag.
	
	The influence of the LTR on the PSD is mainly manifested in the reduction of the power. The simulations of sinusoidal and QPOs signals show that the LTR effect is more important at high frequencies. After the LTR, the signal power decreases monotonically with frequency. For sinusoidal signals, when the frequency is 536 Hz, the peak of the power spectral density is reduced by $50\%$ compared with none LTR time series. For the QPO signals, this value is 458 Hz. We note that the centroid frequency and the FWHM of QPOs does not change in our simulations. So the LTR will not change the Q factor of QPOs. After the LTR, the RMS of QPOs also decreases.
	
	The influence of the LTR on the pulse profile is mainly manifested in the reduction of the pulsed peak, the RMS and the phase shifting of the pulse profile. All of these effects increase monotonically with the decrease of the period. The pulse profile decreases by a half and the phase shifts $\pi$ when the period value is about 2 ms. When the period value is 1 ms, RMS decreases by a half. 
	
	There are two situations for the influence of the LTR on the time lag. The first case is that both light curves are influenced by the LTR. The cross-correlation function is used to calculate the time lag between two signals in the time domain. Simulation results show that the LTR can only reduce the maximum of the cross-correlation function while it does not change the position of it. Therefore, the calculated delay is not changed by the LTR. In the second case, when one signal influenced by the LTR while the other not, a fixed delay of 1.18 ms will be introduced between these two signals, which should be eliminated in the actual calculation. The results in the frequency domain are consistent with those in the the time domain. 
	
	All these results indicate that although the maximum time uncertainty of LE/SCDs is about 1 millisecond, the timing analysis nearing 1 millisecond can also be analysed in some ways, as discussed above. This provides a reference for timing analysis using \emph{Insight}-HXMT/LE. 
	
	\begin{acknowledgements}
		This work is supported by the
		National Key R\&D Program of China (2016YFA0400800)
		and the National Natural Science Foundation of China under
		grants, U1838201, U1838202, U1838101 and  U1938109. This work made use of the data from the Insight-HXMT mission, a project funded by China National Space Administration (CNSA) and the Chinese Academy of Sciences (CAS).  
	\end{acknowledgements}
	
	\bibliographystyle{raa}
	\bibliography{ms2020-0163bib}
\end{document}